\documentclass{elsart3p}
\usepackage{graphicx}
\usepackage{amssymb}

\newcommand{\Ag}{Ag$_5$Pb$_2$O$_6$}

\begin{document}
\begin{frontmatter}
  
  \title{Properties of the Nearly Free Electron Superconductor
    \Ag\ Inferred from Fermi Surface Measurements}

\author[cam]{P. D. A. Mann\corauthref{pdam}}
\author[cam]{M. Sutherland}
\author[cam]{C. Bergemann}
\author[kyoto]{S. Yonezawa}
\author[kyoto,kyoto2]{Y. Maeno}
\address[cam]{Cavendish Laboratory, University of Cambridge,
  J. J. Thomson Avenue, Cambridge, CB3 0HE, United Kingdom} 
\address[kyoto]{Department of Physics, Graduate School of Science,
  Kyoto University, Kyoto 606-8502, Japan} 
\address[kyoto2]{International Innovation Centre,Kyoto University,
  Kyoto 606-8501, Japan} 
\corauth[pdam]{Corresponding author. Tel.: +44 (0)1223 337422; e-mail:
  pdam2@cam.ac.uk} 
\begin{abstract}
  We measured the Fermi surface of the recently discovered
  superconductor \Ag\ via a de Haas-van Alphen rotation study.  Two
  frequency branches were observed and identified with the neck and
  belly orbits of a very simple, nearly free electron Fermi surface.
  We use the observed Fermi surface geometry to quantitatively deduce
  superconducting properties such as the in-plane and out-of-plane
  penetration depths, the coherence length in the clean limit, and the
  critical field; as well as normal state properties such as the
  specific heat and the resistivity anisotropy. \copyright 2006
  Elsevier Science. All rights reserved
\end{abstract}

\end{frontmatter}

\section{Introduction}
\label{sec:intro}

Superconductivity has recently been discovered in \Ag\ 
\cite{yonezawa1}. The material is the first layered silver oxide
superconductor, and one of only a handful of known non-elemental
type-I superconductors. While the normal state of \Ag\ is a good
metal, it exhibits a quasi-$T^2$ dependence of the electrical
resistivity up to room temperature which arises from a yet
unidentified scattering mechanism \cite{yonezawa2}.  There have been
several band structure calculation on this material with contradictory
results \cite{brennan,shein,oguchi}. 

To elucidate these issues, we experimentally determined the Fermi
surface (FS) using the de Haas-van Alphen (dHvA) effect and calculated
the resulting semiclassical values for several superconducting and
normal state parameters.

\section{Experimental Fermi Surface}

\begin{figure}
\centering
\includegraphics[width=3cm]{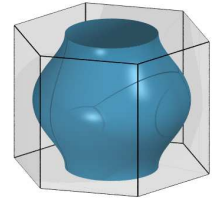}
\caption{\label{fig:FS} The experimentally derived FS of \Ag\ inside
  the first Brillouin zone.}
\end{figure}

Our dHvA experiment, reported in full in reference \cite{sutherland},
revealed that the FS of \Ag\ is nearly free electron-like, as shown
in Fig.~\ref{fig:FS}. Similar to the noble metals, the FS occupies
half the Brillouin zone and punches through its boundary to form neck
orbits. The effective masses for the neck and belly orbits were found
to be $m^\star/m_e = 1.25\pm0.1$ and $1.1\pm0.2$.

The FS is in very good agreement with band structure calculations
by Oguchi \cite{oguchi} and those by Shein and Ivanovskii
\cite{shein}, but differs from those by Brennan and Burdett
\cite{brennan}. However, Oguchi \cite{oguchi} predicts a much lighter
neck orbit ($m^\star/m_e = 0.68$) than we see in experiment, while his
belly orbit mass matches rather well. The only way to reconcile that
calculation with our data is to invoke very anisotropic electron-phonon
coupling over the FS.

\section{Parameterisation and Calculations}

In order to calculate several normal state and superconducting
parameters from our dHvA data, we need to parameterise the
experimental FS and band structure. The lattice parameters are 
$a=5.93$\,\AA\ and $c=6.41$\,\AA, and we obtain a very good fit to our
data (with deviations between the observed and the fitted dHvA
frequencies of 0.1\,kT or less at all angles) by using
\begin{equation}
  \label{eq:loosebinding}
  E = \frac{\hbar^2}{2m^\star} (k_x^2+k_y^2) - 2t_\bot\cos ck_z
\end{equation}
with $m^\star = 1.2 m_e$, $t_\bot = 0.157$\,eV, and a Fermi energy
$E_F = 0.656$\,eV.  While the $c$-axis dispersion formally has a tight
binding form, it should rather be thought of as ``loose binding''
since $t_\bot$ is of order $E_F$. This energy dispersion gives the
same dHvA mass $m^\star = 1.2m_e$ for both neck and belly orbits, in
line with experiment. The Fermi volume in this model gives an electron
density of $n = m^\star E_F/\pi c\hbar^2$ which agrees to within 1\,\%
with the stoichiometric expectation of one electron per unit cell.

\begin{table}
  \caption{\label{table:props}Superconducting and normal state
    parameters of \Ag\ in the limit $T\to 0$. When two values are
    given, they refer to in-plane/$c$-axis values, respectively.}
  \centerline{\begin{tabular}{l@{\ \ }c@{\ \ }c}
    & From FS & Direct \cite{yonezawa1,yonezawa2}
    \\\hline\hline
    $\gamma$ & 3.6\,mJ/K$^2$\,mol(f.u.) & 3.4\,mJ/K$^2$\,mol(f.u.) \\
    $\lambda$ & 81\,nm/116\,nm & ---\\
    $\xi$ & 8.4\,$\mu$m/5.9\,$\mu$m & ---\\
    $B_c$ & 0.33\,mT & 0.2\,mT \\
    $\rho_\bot/\rho_=$ & 2.1 & 6.5 \\\hline\hline
  \end{tabular}}
\end{table}

Various parameters can now be calculated from the band structure as
surface or volume integrals over the Fermi surface. Some, but not all,
of these can be performed analytically. Table~\ref{table:props} lists
the results in the limit $T \to 0$.

First, the specific heat coefficient is proportional to the density of
states at $E_F$; specifically here we get $\gamma = \pi
k_B^2a^2m^\star N_A/2\sqrt{3}\hbar^2$. The penetration depth can be
calculated from the semiclassical Chandrasekhar-Einzel integral
\cite{chandra}, the results are a very familiar in-plane value
$\lambda_=^{-2} = n e^2\mu_0/m^\star$ and a $c$-axis penetration depth
$\lambda_\bot^{-2} = 2 e^2 \mu_0 m^\star c t_\bot^2 / \pi\hbar^4$.
The in-plane and $c$-axis coherence lengths are estimated from the RMS
values of the Fermi velocities via $\xi \simeq 0.18\,\hbar v_F/k_B
T_c$. They are much larger than the typical mean free path, implying
that \Ag\ is almost always in the dirty limit (see also
\cite{yonezawa1}). We nevertheless list the clean limit critical
field in Table~\ref{table:props} also.  The normal state conductivity
tensor can be calculated as a standard Fermi surface integral
\cite{ashcroft}, where we assume a constant mean free path.

\section{Discussion}

The values for $\xi$ and $\lambda$ identify \Ag\ as type-I (in the
clean limit).  The resistivity anisotropy $\rho_\bot/\rho_=$ comes out
smaller than measured \cite{yonezawa2}, indicating either a very
anisotropic scattering rate or uncertainties about the geometry factor
in those measurements. We believe that \Ag\ is the first example of a
nearly free electron superconductor, and it has the ``most spherical''
FS of all known superconductors. We now understand the 
normal state of \Ag\ rather well, but the anomalously strong $T^2$
term in the resistivity remains mysterious.

\section*{Acknowledgments}
We would like to thank J. Fletcher, G. G. Lonzarich, and P. Pyykk\"o
for useful discussions. C.B.\ acknowledges the support of the Royal
Society.

\bibliographystyle{elsart-num}

\end{document}